%% file: main.tex
\documentclass{article}
\usepackage[english]{babel}
\usepackage[utf8]{inputenc}
\usepackage{amsmath,amsthm}
\usepackage{amsfonts}
\usepackage{mathtools}
\usepackage{xcolor}
\usepackage{listings}
\usepackage{ragged2e}
\usepackage{mathtools}

\usepackage{verbatimbox}
\newcommand{\Program}[1]{\smallskip\noindent\textbf{Program \spc: #1}}

\usepackage{enumitem}
\usepackage{array}
\usepackage{hyperref}

\newtheorem{proposition}{Proposition}
\newtheorem{observation}{Observation}

\newtheorem{corollary}{Corollary}
\newtheorem{lemma}{Lemma}

\theoremstyle{definition}
\newtheorem{remark}{Remark}
\newtheorem{definition}{Definition}

\newcommand{\calR}{\mathcal{R}}

\newcommand{\calH}{\mathcal{H}}
\newcommand{\calN}{\mathcal{N}}
\newcommand{\calA}{\mathcal{A}}

\newcommand{\calD}{\mathcal{D}}

\newcommand{\calS}{\mathcal{S}}
\newcommand{\calL}{\mathcal{L}}
\newcommand{\calP}{\mathcal{P}}
\newcommand{\calM}{\mathcal{M}}
\newcommand{\calC}{\mathcal{C}}
\newcommand{\calT}{\mathcal{T}}
\newcommand{\cals}{\mathcal{s}}
\newcommand{\calt}{\textbf{\textit{t}}}
\newcommand{\calv}{\textbf{\textit{v}}}
\newcommand{\calg}{\textbf{\textit{g}}}

\newcommand{\nimrod}[1]{\textcolor{red}{Nimrod says: #1}}
\newcommand{\udi}[1]{\textcolor{blue}{Udi says: #1}}
\newcommand{\gal}[1]{\textcolor{orange}{Gal says: #1}}
\newcommand{\ouri}[1]{\textcolor{magenta}{Ouri says: #1}}

\newcounter{pc}
\newcommand\spc{\addtocounter{pc}{1}\thepc}

\newcommand{\hidden}[1]{}

\title{Digital Social Contracts: A Foundation for an Egalitarian and Just Digital Society}

\author{Luca Cardelli \\
University of Oxford \\
\url{luca.a.cardelli@gmail.com}
\and
Liav Orgad \\
European University Institute \\
\url{liav.orgad@eui.eu}
\and
Gal Shahaf \\
Weizmann Institute \\
\url{gal.shahaf@weizmann.ac.il}
\and
Ehud Shapiro \\
Weizmann Institute \\
\url{ehud.shapiro@weizmann.ac.il}
\and
Nimrod Talmon \\
Ben-Gurion University \\
\url{talmonn@bgu.ac.il}}

\date{}

\begin{document}

\maketitle

\pagestyle{plain}

\begin{abstract}
Almost two centuries ago Pierre-Joseph Proudhon proposed social contracts --  voluntary agreements among free people -- as a foundation from which an egalitarian and just society can emerge.  A \emph{digital social contract} is the novel incarnation of this concept for the digital age:  a voluntary agreement between people that is specified, undertaken, and fulfilled in the digital realm. It embodies the notion of ``code-is-law'' in its purest form, in that a digital social contract is in fact a program -- code in a social contracts programming language, which specifies the digital actions parties to the social contract may take; and the parties to the contract are entrusted, equally, with the task of ensuring that each party abides by the contract. Parties to a social contract are identified via their public keys, and the one and only type of action a party to a digital social contract may take is a ``digital speech act'' -- signing an utterance with her private key and sending it to the other parties to the contract.

Here, we present a formal definition of a digital social contract as agents that communicate asynchronously via digital speech acts, where the output of each agent is the input of all the other agents. 
We outline an abstract design for a social contracts programming language and show, via programming examples, that key application areas, including social community; simple sharing-economy applications; egalitarian currency networks; and democratic community governance, can all be expressed elegantly and efficiently as digital social contracts.   Possible extensions, described in companion papers, include autonomous deterministic agents akin to ``smart contracts'', and joint agents executed jointly by several parties to the contract; a definition of a distributed implementation of digital social contracts in the presence of faulty agents; and egalitarian and just currency networks, suitable for realization via a network of digital social contracts.
\end{abstract}

\section{Introduction}
Over the last three decades, the Internet has mushroomed from 0 to 4.6 billion active “users”, 60 per cent of the world population (and more than 90 per cent in the developed world), the fastest diffusion in human history. But what kind of “society” has it created? The digital realm includes a few powerful entities, who control the entire space, and billions of “data subjects”, as termed by the EU’s General Data Protection Regulation (GDPR), who have almost no rights and cannot vote, be consulted, or influence decision-making. They have no due process (e.g., in censorship, account blocking, or privacy violation) and are not the owners of their data. The structure of the digital space remains “feudal” in nature—people are not even perceived as digital “citizens” but as “users”—and it is not based on the ideals on which citizenship in democracies is grounded: liberty, equality, and popular sovereignty. The digital space is further rampant with sybils (fake/duplicate accounts) and bots, making it a hostile environment and untenable for democratic governance. To a large extent, digital technology is the great unequalizer in today’s world, making a few digital “barons” rich and powerful by profiting from the multitudes performing digital labour without compensation. Like feudal lords of yore, the digital barons set their platform’s rules of conduct (legislative), interpretation (judicial), and implementation (executive), and violate the dictum “no taxation without representation”. New technologies do not support the ideals of democracy. Rather, the design of the digital world presents a dystopian future, with the global reach of “surveillance capitalism” \cite{zuboff2019surveillance} robbing citizens of both power and wealth. The structure of the digital world resembles the “state of nature” (i.e., a social disorder).

In order to establish a \textit{self-governed}, \textit{egalitarian}, and \textit{just} \textit{society} in the digital world, this article advocates the development of digital social contracts to govern digital human behavior. A digital social contract is the digital counterpart of social contract theories, as envisioned centuries ago by the founding philosophers of the Enlightenment – Thomas Hobbes, John Locke, Jean-Jacques Rousseau, and Pierre-Joseph Proudhon – and more recent thinkers, such as John Rawls \cite{rawls2009theory,  rousseau2002social, proudhon2004general, locke2014second, hobbes1914leviathan}. Social contract theories are the cornerstone of modern democracies (they explain the conditions under which people rationally agree to submit their rights to a legitimate authority in return for collective benefits, such as protection, freedom, or justice) yet, thus far, have remained merely an abstract ideal, based on a hypothetical narrative and the presupposition of tacit, rather than explicit, consent \cite{riley2013will}. A digital social contract offers, for the first time  in history, a conceptual model to translate the social contract from a hypothetical theory into a political reality, by utilizing technology in the service of the most important ideas of the Enlightenment. In short, a digital social contract is a voluntary agreement between individuals, specified, undertaken, and fulfilled in the digital realm. It allows for conditions to reach the social contract to be programmed by generating equal access to information. It further enables the programming of the method of agreement by implementing a “one person, one vote” system. And it provides the opportunity to check contractual outcomes in ways that comply with requirements of fairness, for instance, by limiting unequal distributional outcomes. We aim for digital social contracts that are equal (in sharing power), just (in allocating resources), and self-governed (among the participating members ). Hence, the equal sharing of power would require mitigating sybils (fake and duplicate identities) and launching digital currencies where distributed and egalitarian coin minting provides a form of Universal Basic Income~\cite{van2017basic} to members.

On the technological design, digital social contracts are programs for a distributed, locally-replicated, asynchronous blockchain architecture. Digital social contracts embody the notion of “code-is-law” in that they are a code in a social-contract programming language, which specifies the digital actions parties may take and the terms. Unlike present digital platforms, this code is not inflicted on the user by a monopolistic digital entity to its financial benefit; rather, people who trust each other enter into a code-regulated relationship voluntarily. Digital social contracts allow for cooperative behavior because they disallow faulty, non-cooperative actions; a party can behave only according to the contract. Agents are expected to employ genuine identifiers and function as both actors and validators. An agent may participate simultaneously in multiple social contracts; similarly, each contract may have a different set of agents as parties; the result is a hypergraph with agents as vertices and contracts as hyperedges. Signed indexed transactions (“digital speech acts”), specified and carried out according to a contract, are replicated only among parties to a contract, who ratify them, and are synchronized independently by each agent. A transaction carried out is finalized when ratified by an appropriate supermajority of the parties to the contract. A party to a contract is typically an agent who embodies a natural person via her genuine identifier, but, if deterministic, could also be a synthetic agent (aka ``legal person''), which functions algorithmically, very similar to a “smart contract” of standard blockchains \cite{reijers2016governance}.

In 2017, Mark Zuckerberg published a vision on how to turn Facebook from a social network to a “Global Community” governing the planet in most aspects of life, where every user can vote on “global problems that span national boundaries” and “participate in collective decision-making” \cite{zuckerberg2017building}. According to Zuckerberg, “Facebook can explore examples of how community governance might work at scale” yet, in his global community, people have no rights and bear no responsibilities; they are a means to make profit for Facebook's stakeholders. Zuckerberg’s community turns the problem on its head – Facebook is presented as the solution, rather than the problem. The starting point of this article is radically different than Facebook and similar-minded companies in two aspects: 1) vision: it starts from the premise that there can and should a digital public life, complementary to the traditional public life in the physical world, by which political communities govern their life; 2) technology: The only technology that provides sovereignty to its community of users is blockchain. Nobody can unplug Bitcoin or Ethereum; both will keep running as long as someone somewhere keeps running their protocols. Alas, current blockchain technologies are environmentally harmful and plutocratic – they are neither egalitarian nor just as sovereignty is shared not equally but according to one’s wealth, i.e., one’s ability to prove work or stake.  We aim to offer an alternative.
In summary, \emph{digital social contracts} are:
\begin{enumerate}
    \item \textbf{Conceptually}, a voluntary agreement among individuals, specified, undertaken, and fulfilled in the digital realm.
    \item \textbf{Mathematically}, an abstract model of computation – a transition system specifying concurrent, non-deterministic asynchronous agents engaged in digital speech acts. 
    \item \textbf{Computationally}, a program in a Social-Contracts Programming Language with which social contracts among computational agents are easy to express, comprehend and debug. 
    \item \textbf{Technologically}, a programmable, distributed, locally-replicated, asynchronous blockchain architecture, operating on mobile phones, tagged by the genuine identifiers of their owners. 
\end{enumerate}

Two key concepts are employed in digital social contracts:
\begin{enumerate}
    \item \textbf{Digital Speech Acts}: Each party to a digital social contract is equipped with a key pair -- a public key and a private key (PKI), with which they can digitally sign text strings and verify each other's signatures. All that a computational agent can do, as a party to a digital social contract, is perform and perceive digital speech acts, namely sign a digital utterance and send it to the other agents that are parties to the contract, or receive signed digital utterances from these agents.
    \item \textbf{The Person as an Oracle}: Computational agents in a digital social contract face nondeterministic choices: Which room to book, when and where?  How much to pay and to whom?  How to vote on a proposal to accept a new member to the contract? Computational agents have no volition or free will and hence cannot make such choices on their own. Therefore, nondeterministic choices in the social contract code are interpreted as Oracle calls, where the human operating the computational agent serves as the Oracle. In other words, nondeterministic choices in the code of an agent specify the interactions between the ``software'' and the ``user'', namely between the computational agent and the person operating it:  When faced with a nondeterministic choice, the computational agent consults its human operator as to which choice to make.
\end{enumerate}

\hidden{
\input{groupcontract.tex}

\subsection{Community: Example of a Digital Social Contract}

A simple, useful example of a digital social contract is a social community.
Existing solutions for community communication rely on servers controlled by the service provider,  which undermine the sovereignty of the community, let alone its privacy.
The \textbf{community} digital social contract has roles and actions as detailed in Table~\ref{figure:community}.  It specifies a private social community, of which the manager is the sovereign; anyone can accept, reject or ignore an invitation to become a member in a community.  Turning this autocratic digital social contract into an egalitarian one is discussed herein.
}



\subsection{Related Work}\label{section:relatedwork}

The concepts and design presented here are reminiscent of the notions of blockchains~\cite{blockchainbook}, smart contracts~\cite{smartcontracts}, and their programming languages~\cite{solidity}. Hand-in-hand with these we are working on egalitarian currency networks~\cite{EgalitarianCurrencyNetworks}, an egalitarian and  just alternative to existing plutocratic cryptocurrencies such as Bitcoin~\cite{bitcoin} and Ethereum~\cite{solidity}.

A fundamental tenet of our design is that social contracts are made between people who know and trust each other,  directly or indirectly via other people that they know and trust. This is in stark contrast to the design of cryptocurrencies and their associated smart contracts, which are made between anonymous and trustless accounts.  A challenge cryptocurrencies address is how to achieve consensus in the absence of trust, and their solution is based on proof-of-work~\cite{POW} or, more recently, proof-of-stake~\cite{POS} protocols. In contrast, social contracts are between known and trustworthy individuals, each expected to posses a genuine (unique and singular) identifier~\cite{ggid} (see therein discussion on how this can be ensured).  Hence, a different approach can be taken.  In our approach, the integrity of the ledger of actions taken by the parties to the social contract is preserved internally, among the parties to the agreement, not between external anonymous ``miners'', as in cryptocurrencies.  This gives rise to a much simpler approach to fault tolerance, as discussed in the companion paper~\cite{ftdsc}.

\hidden{
\subsection{Paper Structure}

Next we describe a formal, descriptive model for digital social contracts, as well as a possible design for a programming language to program social contracts in Section~\ref{section:model}.  
We then outline several examples of social contracts in Section~\ref{section: examples}; specifically, we describe a general design using the programming language for three broad settings: social networks, shared economy and sovereign currencies. We then discuss the realization of egalitarian social contracts and democratic governance in Section~\ref{section:democracy}, and conclude with intriguing questions for further research.
}

\section{Digital Social Contracts}\label{section:model}

Here we describe a formal model for digital social contracts.

\subsection{Preliminaries}

We assume a given finite set of agents $V$, each associated with a \emph{genuine} (unique and singular)~\cite{ggid} identifier, which is also a public key of a key-pair.\footnote{We identify the set of agents $V$ with the set of parties to the agreement.  Extensions will allow an agent to be a party to multiple agreements, and different agreements to have different sets of agents as parties.}
We expect agents to be realized by computer programs operating on personal devices (e.g. smartphones) of people.  Hence, we refer to agents as ``it'' rather than as he or she.

We identify an agent $v \in V$ with its genuine public identifier, and  denote by $v(s)$ the result of agent $v$ signing the string $s \in \calS$ with the private key corresponding to $v$. We assume computational hardness of the public-key system, namely that signatures of an agent with a given identifier cannot be produced without possessing the private key corresponding to this identifier.
To avoid notational clutter we do not distinguish between finite mathematical entities and their string representation. Identifying agents with their genuine identifiers makes $V$ totally ordered (by the numeric value of the identifier, namely the public key)  and hence allows defining tuples and Cartesian products indexed by $V$. If $t$ is a tuple indexed by $V$, then we use $t[v]$ to refer to the $v^{th}$ element of $t$.   We say that $t' = t$, except for $t'[v]:= x$, to mean that the tuple $t'$ is obtained from $t$ by replacing its $v^{th}$ element by $x$.  In particular, if $t[v]$ is a sequence, then we say that $t' = t$, except for $t'[v] := t[v]\cdot x$, to mean that $t'$ is obtained from $t$ by appending $x$ to the sequence $t[v]$.  

All agents are assumed to be connected via a reliable asynchronous communication network (without assuming a known time limit on message arrival), and require all messages to be authenticated.
Informally, the only things an agent can do as a party to a digital social contract are  (i) perform a \emph{digital speech act}~\cite{ggid}, defined next; (ii) observe digital speech acts performed by others; and (iii) change internal state.\footnote{While the formal definition allows a digital speech act to employ an arbitrary string, its intended use is to take meaningful actions.  As parties to a social contract employ strings that are meaningful to the other parties, we believe we do not conjure ``speech acts'' in vain.}

A key characteristic of a speech act taken in the physical world is that all people present indeed perceive the person taking the act as well as the act itself. We capture this characteristic by requiring that a digital speech act be (i) digitally signed by the person taking it

\begin{definition}[Digital Speech Act, $v$-act]\label{definition:csa}
Given a set of agents $V$, a \emph{digital speech act} of agent $v \in V$ consists of (i) signing an utterance (text string) $s$,  resulting in $m=v(s)$; and (ii) broadcasting the message $m$ to $V$. We refer to a digital speech act by $v$ resulting in the  signed action $m$ as the $v$\emph{-act} $m$, and let $\calM$ be the set of all $v$-acts for all $v \in V$.
\end{definition}

We employ a standard notion of a transition system:
\begin{definition}[Transition System]
A \emph{transition system} $TS=(S,s_0,T)$ consists of a set of states $S$, an initial state $s_0 \in S$, and a set of transitions $T$,  $T \subseteq S \times S$, with $(s,s') \in T$ written as $s \xrightarrow{} s'$.  The set $s \rightarrow * = \{ \hat s~ |~ s \xrightarrow{} \hat s \in T\}$ is the \emph{outgoing transitions} of $s$. A \emph{run} of $TS$ is a sequence of transitions $r = s_0 \xrightarrow{} s_1 \xrightarrow{}  \ldots $ from the initial state. A \emph{family} of transition systems $\calT(S)$ over $S$ is a set of transition systems of the form $(T,S)$ where $T$ is a set of transitions over $S$.
\end{definition}

A family of transition systems $\calT(S)$ is best thought of as the abstract, syntax-free counterpart of a programming language, where each transition system $(T,S) \in \calT(S)$ is the abstract counterpart of a program in the language.

\begin{definition}[Implementation]\label{definition:implementation}
Assume two transition systems $TS=(S,s_0,T)$ and $TS'=(S',s'_0,T')$,
and a function  $f: S' \rightarrow S$, where $f(s'_0) = s_0$, referred to as a \emph{semantic mapping}.
Then  $TS'$ \emph{implements} $TS$ via $f$ if:
\begin{enumerate}
    \item For every $TS$ transition $s_1\xrightarrow{} s_2 \in T$, there is a sequence of $TS'$ transitions $s'_1\xrightarrow{}\ldots \xrightarrow{} s'_n  \in T'$, $n\ge2$, such that $f(s'_1)=s_1$ and $f(s'_n)=s_2$.
    
    \item For every $TS'$ transition $s'_1\xrightarrow{} s'_2 \in T'$, either $f(s'_1) = f(s'_2)$ or  $f(s'_1) \xrightarrow{} f(s'_2) \in T$.
\end{enumerate}
$TS'$ \emph{can implement} $TS$ if there is a semantic mapping $f: S' \rightarrow S$ via which $TS'$ implements $TS$.\qed
\end{definition}

\begin{definition}[Morphism, Strict]
Given two transition systems $TS=(S,s_0,T)$ and $TS'=(S',s'_0,T')$, a semantic function  $f: S' \rightarrow S$, where $f(s'_0) = s_0$ is a \emph{morphism} if $s \xrightarrow{} \hat s \in T$ implies $f(s) \xrightarrow{} f(\hat s) \in T'$; and \emph{strict} if $f(s) \xrightarrow{} f(\hat s) \in T'$ implies $s \xrightarrow{} \hat s \in T$.
\end{definition}  
That is, in a strict morphism the implementing process provides all the transitions of the implemented process \cite{hesselink1988deadlock}.

A strict morphism is an implementation in which every transition of the implementing transition system is mapped to exactly one transition of the implemented transition system, and vice versa.  Hence:
\begin{observation}
A strict morphism among two transition systems is an implementation.
\end{observation}

\begin{definition}[Compiler]\label{definition:compiler}
Given two families of transition systems, $\calT\calS$, $\calT'(S')$, consider a pair of functions $(f,c)$, where  $f: S' \rightarrow S$ is a semantic mapping and $c: \calT(S) \xrightarrow{} \calT'(\calS')$ is  referred to as a \emph{compiler}.  Then $(f,c)$   \emph{implement} $\calT(S)$ if
for each $TS \in \calT(S)$, $c(TS)$  implements  $TS$ via $f$.  We say that $\calT'(S')$ \emph{can implement} $\calT(S)$ if there is a pair of functions $(f,c)$ as above that implement $\calT(S)$. 
\end{definition}

Eventually, we aim to use the framework developed here to indeed prove that the Social-Contracts Programming Language we are designing implements the abstract transition system of social contracts described herein.

\subsection{A Digital Social Contract Transition System}

Here we define digital social contracts in the abstract; in particular, we do not address a distributed realization nor agent faults.  These issues are addressed in a companion paper~\cite{ftdsc}.

\begin{remark}[Parties, Roles, and Parameterized Social Contracts]
Agents should be able to participate simultaneously in multiple social contracts.  Furthermore, a contract may have multiple parties with the exact same role.  To address this properly, we should describe digital social contracts as a set of roles, each specified by a parameterized procedure, and bind the formal parameters to actual agent identifiers upon execution of the contract. This is akin to the standard legal practice of using a textual contract template, with role names as parameters  (e.g. Landlord, Tenant), and filling in the identities of the  parties assuming these roles in an instance of this contract template upon its signature. We defer this distinction between formal and actual parameters in a social contract to avoid notational clutter, and name the parties to the social contract by their genuine identifiers, i.e., their public keys. Later we describe a design for a practical Social-Contracts Programming Language, which will naturally make this distinction.
\end{remark}

A digital social contract consists of a set of agents, identified via public keys,  and connected via a reliable, asynchronous, order-preserving communication network. Namely, we assume that between any two agents in the network, the network delivers all messages sent from one agent to another -- correctly, eventually, and in the order sent.\footnote{As agents have public keys with which they sign their sequentially-indexed messages, such a network can be easily realized on an asynchronous network that is not order-preserving and is only intermittently reliable.}
All that agents can do within a digital social contract is perform digital speech acts (henceforth, \emph{acts}), which are sequentially-indexed utterances, signed by the agent and sent as messages to all other agents that are parties to the contract, as well as receive such messages. Example acts are ``I hereby transfer three blue coins to Sue'' and ``I hereby confirm the room reservation by Joe''.  A digital social contract specifies the digital speech acts each party may take at any state. For example, a social contract regarding the blue currency may specify that I can say that I transfer three blue coins to Sue only in a state in which I in fact have at least three blue coins; and I can say that I confirm Joe's room reservation only if I have not already confirmed a conflicting reservation.

Our abstract notion of a digital social contract identifies a digital social contract with a transition system. 
The state of a digital social contract, referred to as a \emph{ledger}, is composed of the states of each of the agents participating in the contract, referred to as their \emph{history}, which is but a sequence of digital speech acts.  The history of an agent $v$ consist of digital speech acts it experienced, in particular: acts by $v$, referred to as \emph{$v$-acts}, in the order taken, interleaved with acts by the other agents, in the order received from the network. Hence, at any ledger that is a state of the computation, the $u$-acts in the history of agent $v$ must be a prefix of the $u$-acts in the history of $u$, for any $u$ and $v$. We call such a ledger \emph{sound}.   Next we formalizes this informal description.

We assume a given set of actions $A$. Signing an action by an agent $v$ makes the action non-repudiated by $v$. All that an agent that is party to a digital social contract does, then, is perform digital speech acts, as well as receive such acts performed by others, resulting in a sequence of non-repudiated acts - a history of crypto speech acts.

\hidden{
\gal{To DSC2:
Actions performed by an agent are first sequentially-indexed and then signed. 
Signing sequentially-indexed actions also makes the order of the actions of $v$ non-repudiated by $v$.}     
}

\begin{definition}[$v$-act, History]
%
We refer to $m=v(a)$, the result of signing an  action $a \in A$ by agent $v \in V$, as a $v$\emph{-act}, let $\calM$ denote the set of all $v$-acts by all $v \in V$, and refer to members of $\calM$ as \emph{acts}.
A \emph{history} is a finite sequence of acts $h = m_0, m_1, \ldots m_n$, $n \in \calN$,  $m_i =v_i(a_i) \in \calM$, $i\in [n]$,  $a_i \in A$, $v_i \in V$. 
The set of all histories is denoted by $\calH$.
\end{definition}

Next, we define a pair histories to be consistent if they agree on the subsequences signed by each agent.

\begin{definition}[Prefix, Consistent Histories]\label{definition:consistent histories}
Given a sequence $s= x_1,x_2, \ldots x_n$, a sequence $s'$ is  a \emph{prefix} of $s$, $s' \preceq s$, if $s'= x_1,x_2, \ldots x_k$, for some $k \le n$. We apply the notation $h[u]$, of restricting a history $h$ to the subsequence of  $u$-acts. 

Two agent histories $h,h'\in \calH$ are \emph{consistent} if either $h[v] \preceq h'[v]$ or vice versa for every $v \in V$. 
\end{definition}

\hidden{
\gal{I'm trying to clean this def. This is the original:
\begin{definition}[Consistent Histories, Ledger Diagonal and Prefix]\label{definition:consistent diagonal prefix}
Given a ledger $l \in \calL$, two agent histories $l_u, l_v$, for $u, v \in V$, are \emph{consistent} if $l_u[v] \preceq l_v[v]$, $l_v[u] \preceq l_u[u]$, and $l_u[w] \preceq l_v[w]$ or vice versa for every $w \ne u, v \in V$. A ledger is \emph{consistent} if $l_u$ and $l_v$ are consistent for every $u,v \in V$. Given a ledger $l$,  the \emph{diagonal} of $l$,  $l^*$, is defined by $l^*_v := l_v[v]$ for all $v \in V$.
Given two ledgers $k, l \in \calL$, define $k \preceq l$ if $k^*_{v} \preceq l^*_{v}$ for every $v \in V$. 
\end{definition}
}
}

A ledger of a digital social contract is an indexed tuple of  histories of the parties to the contract. 
Ledgers are the states of the social contract transition system.  

\begin{definition}[Ledger]\label{definition:ledger}
A \emph{ledger} $l \in \calL := \calH^V$ is a tuple of histories indexed by the agents $V$, where $l_v$ is referred to as the \emph{history of agent $v$ in $l$}, or as \emph{$l$'s $v$-history}.
\end{definition}

Following Definition \ref{definition:consistent histories}, we apply the notation $l_v[u]$ to denote the subsequence of $u$-acts in the $v$-history in $l$. 

Our key assumption is that in a ledger that is a state of a computation, each agent's history is \emph{up-to-date about its own acts}. Therefore, for agent histories to be consistent with each other, the history of each agent can include at most a prefix (empty, partial or complete) of every other agent's acts.  In particular , the history of $v$ cannot include $u$-acts different from the one's taken by $u$; nor can it run ahead of $u$ and ``guess''  acts $u$ might take in the future. 

Note that a ledger is not a linear data structure, like a blockchain, but a tuple of independent agent histories, each of them being a linear sequence of acts. Furthermore, note that $l_v$, the history of $v$ in ledger $l$, contains, of course, all $v$-acts, but it also contains acts by other agents received by $v$ via the communication network.   Also, note that the $v$-history $l_v$ may be behind on the acts of other agents but is up-to-date, or even can be thought of as the authoritative source in $l$, regarding its own $v$-acts. 

\begin{definition}[Ledger Diagonal, Sound Ledger]\label{definition:consistent diagonal prefix}
Given a ledger $l \in \calL$, the \emph{diagonal} of $l$,  $l^*$, is defined by $l^*_v := l_v[v]$ for all $v \in V$. A ledger is \emph{sound} if $l_u[v] \preceq l^*_v$ for every $u,v \in V$. 
\end{definition}

\hidden{
\gal{Perhaps "sound ledger" is a better term then "consistent".}

\gal{One ledger is a prefix of a second if the second has the same acts as the first, and possibly some more, as evidenced by their diagonals.
\begin{definition}[Ledger Prefix]\label{definition:consistent diagonal prefix}
Given two ledgers $k, l \in \calL$, define $k \preceq l$ if $k^*_{v} \preceq l^*_{v}$ for every $v \in V$. 
\end{definition}
Do we actually use this def of ledger prefix? If not, I suggest dismissing it.}
}

As we assume that each agent is up-to-date about its own acts,  the diagonal contains the complete sequence of $v$-acts for each agent $v \in V$. Thus, each agent history in a ledger reflects the agent's ``subjective view'' of what actions were taken, where the diagonal of a ledger contains the ``objective truth'' about all agents, as the following observation shows. 

\hidden{
\gal{This is the original observation, which I believe not to be an observation but a replication of the Definition of consistent ledger:
\begin{observation}[Consistency and the Diagonal]
A ledger $l \in \calL$ is consistent if and only if $l_v[u] \preceq l^*_u$  for every $v, u \in V$.
\end{observation}
\begin{proof}
Assume that  $l_u, l_v$ are pairwise-inconsistent for some $u, v \in V$. Then, according to the definition of pairwise consistency, either $l_u$ and $l_v$  disagree on the $w$-acts of a third agent $w\ne u, v \in V$, in which case $l_v[w] \not\preceq l^*_w$ or $l_u[w] \not\preceq l^*_w$.  Or they disagree with each other, in which case  $l_v[u] \not\preceq l_u[u] = l^*_u$, 
or $l_u[v] \not\preceq l_v[v] = l^*_v$.  In each of these cases, $l_v[u] \not\preceq l^*_u$  for some $v, u \in V$, which proves the ``if'' direction.
Assume its not the case that $l_v[u] \preceq l^*_u$  for every $v, u \in V$.  As $l^*_v = l_v[v]$ by definition, let $u\ne v$ be agents for which $l_v[u] \not\preceq l^*_u$.  But $l_u[u] = l^*_u$ by definition, hence $l_v[u] \not\preceq l_u[u]$, hence $l$ is not pairwise consistent, which proves the ``only if'' direction.
\end{proof}
Instead, I suggest the following:
}
}

\begin{observation}
    If a ledger $l$ is sound, then every pair of agent histories in $l$ is consistent.
\end{observation}

\hidden{
\begin{figure}
    \centering
    \caption{Agent States, Transitions and Programs: A $v$-act $m_i$, a  $v$-history, a ledger $l$, an input transition (blue) inputs the $u$-act $m$, adding it to the history $l[u]$ of $u$ in $l$;  an output transition (red) that outputs the $v$-act $m'$, adding it to the history $l[v]$ of $v$ in $l$, resulting in $l'$, changes the ledger of $v$ from $l$ (to left of grey line) to $l'$ (to left of blue or red line, respectively). The \textbf{Input} transition followed by the \textbf{Output} transition can be abbreviated in a social contract program by a rule $S, m \xrightarrow{} m', S'$, in which $S$ summarizes $l$ and $S'$ summarizes $l'$ (blue and red combined).}
    \label{figure:elements}
\end{figure}
}

We have defined the states of the transition system, and are now ready to define the transitions themselves.  

\begin{definition}[Transitions]\label{definition:ledger-transition}
The set of \emph{social contract transitions} $\calT \subseteq  \calL \times \calL$ consists of all pairs
$t= l \xrightarrow{} l' \in \calL \times \calL$ where
$l' = l$ except for $l'_v := l_v \cdot m$, $m=u(a) \in \calM$ for some $u,v \in V$ and $a \in A$.  The transition $t$ is referred to as a $v$-transition, and can also be written as $t= l \xrightarrow{(v,m)} l'$.\qed
\end{definition}

\begin{definition}[Input, Output and Sound Transitions]
    A $v$-transition $t= l \xrightarrow{(v,u(a))} l'$ is \emph{output} if $u=v$ and \emph{input} if $u\ne v$. A transition is \textit{sound} if it is either input with $l'_v[u] \preceq l_u[u]$ or output. 
\end{definition}

\begin{observation}[]
If $l$ is sound and $l \xrightarrow{} \hat{l} \in \calT$ is sound then $\hat{l}$ is sound.
\end{observation}
\begin{proof}
An Output $v$-transition does not violate soundness as $\hat{l}_v[v] = \hat{l}^*_v$ by definition.
An Input $v$-transition does not violate soundness thanks to the requirement that $\hat{l}_v[u] \preceq l_u[u] = \hat{l}_u[u] = \hat{l}^*_u$.
\end{proof}

The next definition aims to capture formally the following two informal requirements:  
\begin{enumerate}
    \item Output: That an agent $v$ may take a $v$-act in state $l$ based solely on its own history $l_v$.  In particular, $v$ is free to ignore, or to not know, the other agents' histories in $l$.
    \item Input: That $v$ must accept messages from the communication network in any proper order the network chooses to deliver them. In particular, in any state $l$, $v$ can accept from any other agent $u$ the next consecutive $u$-act~$m$ not yet in $v$'s history.
\end{enumerate} 

\begin{definition}[Closed Set of Transitions]
A set of transitions $T \subseteq \calT$ is:
\begin{enumerate}
    \item \emph{Output-closed} if for any  $v$-act $m$, $v \in V$, $m \in \calM$, and any two output transitions $t = l \xrightarrow{(v,m)} \hat{l}$, $t' = l' \xrightarrow{(v,m)} \hat{l}' \in \calT$,  if $l_v = l'_v$ then $t \in T$ iff $t' \in T$. Namely, if every output $v$-transition is a function of $l_v$, independently of $l_u$ for all $u\neq v$.
  \hidden{  
    \gal{I believe "Output-independent" is a better term then "Output-closed"}}
    \item \emph{Input-closed} if $T$ contains every sound input transition in $\calT$.
    \item  \emph{Closed} if it is output-closed and input-closed.
\end{enumerate}
\end{definition}

\begin{definition}[Digital Social Contract]\label{definition:SC}
A \emph{digital social contract} among a set of agents $V$ is a transition system  $SC =(S,l_0,T)$ with ledgers as states $S \subseteq \calL$, initial state $l_0 := \Lambda^V$, and a closed set of transitions $T \subseteq \calT$. The family of all digital social contracts over $S$ is denoted by $\calS\calC(S)$, and $\calS\calC(\calL)$ is abbreviated as $\calS\calC$.
\end{definition}

\hidden{
See Figure~\ref{figure:ftdscts1}.
}
We refer to $\calS\calC$ as the family of \emph{abstract social contracts}, in contrast to its more concrete implementations discussed here and elsewhere.  In particular, we view abstract social contracts as specifications for programs in the Social-Contracts Programming Language, discussed next.

\hidden{
\begin{figure*}
    \centering
    \includegraphics[scale=0.3]{FIGURES/FTDSCTS1}
    \caption{Four states of a transition system (i.e., a digital social contract). Each circle denotes a state, where the specific state is represented in the table; each row and column corresponds to an agent (we have $3$ agents, $u$, $v$, and $w$) such that, e.g., the cell $(v, w)$ is $v$'s view of $w$'s history at this state (the history is implicitly indexed from the top of the cell to its bottom). Transitions are depicted as directed arcs: $t2$ corresponds to \textbf{Output}$(w)$-transition (as $w$ outputs $w2$); while $t1$ and $t3$ both correspond to an \textbf{Output}$(u)$-transition (as $u$ outputs the message $u3$. Note that monotonicity implies that, since $t1$ is present, then $t3$ must be present as well (that is, as $u$ is able to output $u3$ following $t1$ then $u$ is also able to output $u3$ following $t3$, as the state on the left is a prefix of the state on the bottom.} 
    \label{figure:ftdscts1}
\end{figure*}
}

\hidden{
\begin{definition}[Agent Liveness]
A run $r= l_0 \xrightarrow{} l_1 \xrightarrow{} \cdots$ of a digital social contract $(S,l_0,T)$ is \emph{$v$-live} for $v \in V$
if for every $i\ge 0$ for which there is a $v$-transition from state $l_i$, there is some $j \ge i$ for which a $v$-transition is taken from state $l_j$, namely $l_j \xrightarrow{} l_{j+1}$ is a $v$-transition. A run $r$
satisfies \emph{agent liveness} if it is $v$-live for every $v \in V$.
\end{definition}
}

This completes the definition of the abstract computational model of digital social contracts. 
Liveness and fairness requirements for digital social contracts will be discussed in subsequent work.

\begin{remark}
Note that in the present asynchronous model, neither agent histories nor ledgers have a built-in notion of time.  We would like, however, social contracts to be able to have a notion of time and refer to it.
Adding time to digital social contracts is an anticipated future extension.  
\end{remark}

\hidden{
\gal{Consider the following additions:}

\subsection{Taxonomy of Social Contracts}

A social contract $SC =(S,l_0,T)$ is said to be:
\begin{itemize}
    \item A \textit{fixation}, if $(l_0 \rightarrow l) \notin T$ for all $l\in \calL$ (in particular, if $T=\emptyset$).
    \item An \textit{Anarchy}, if $T$ is the set of all sound transitions.
    \item An \textit{oligarchy}, if there exists a subset $V'\subset V$ such that for every $v\in V\setminus V'$ all input $v$-transitions are not in $T$.
    \item A \textit{dictatorship}, if it is an oligarchy with $|V'|=1$.
    \item \textit{Egalitarian}, if ...
\end{itemize}
}

\subsection{A Currency Community: Example of a Digital Social Contract}

Let $A = \{pay(v)\}_{v\in V}$ be the set of acts. A message $m=u(pay(v))$ corresponds to a payment of 1 coin from $u$ to $v$. A transition $t = l \xrightarrow{w,u(pay(v))} l'$ records in $l_w$ a payment of 1 coin from $u$ to $v$. Agent $u$ can record a payment from $u$ to $v$ only if the balance of $u$, according to $l_u$, is positive. 

Formally, let $l_u = m_1,m_2,...,m_n$, with $m_i=w_i(pay(v_i))$. Set $bal(u):= c + |\{1\leq i\leq n:\>\> v_i = u \}| - |\{1\leq i\leq n:\>\> w_i = u \}|$, for some $c>0$, and let $T = \{l \xrightarrow{u,u(pay(v))} l': \>\> bal(u) > 0\}$. Let $T\subseteq \tilde{T}$ denote the closure of $T$ in $\calT$.

The \emph{digital social contract} $SC =(S,l_0,\tilde{T})$ over the set of agents $V$ specifies a currency community where each agent $u\in V$ is granted an initial endowment of $c$ coins.

Anticipating the Social Contracts Programming Language defined below, here is an SCPL code defining the role of an agent in this community, with an initial endowment of 10 coins:

\Program{Currency with an Initial Endowment}
\begin{verbatim}
agent --> agent(10).
    
agent(Balance), Other(pay(Self)) -->
    agent(Balance') where Balance' := Balance + 1.
   
agent(Balance) --> 
    pay(Other), agent(Balance') 
    where Balance > 0 & Balance' := Balance - 1.
\end{verbatim}
Note that the program is nondeterministic in that an agent may choose nondeterministically any \verb|Other| to pay to.  This notion is discussed at length below.

\hidden{
\begin{figure}
    \centering
    \includegraphics[scale=0.35]{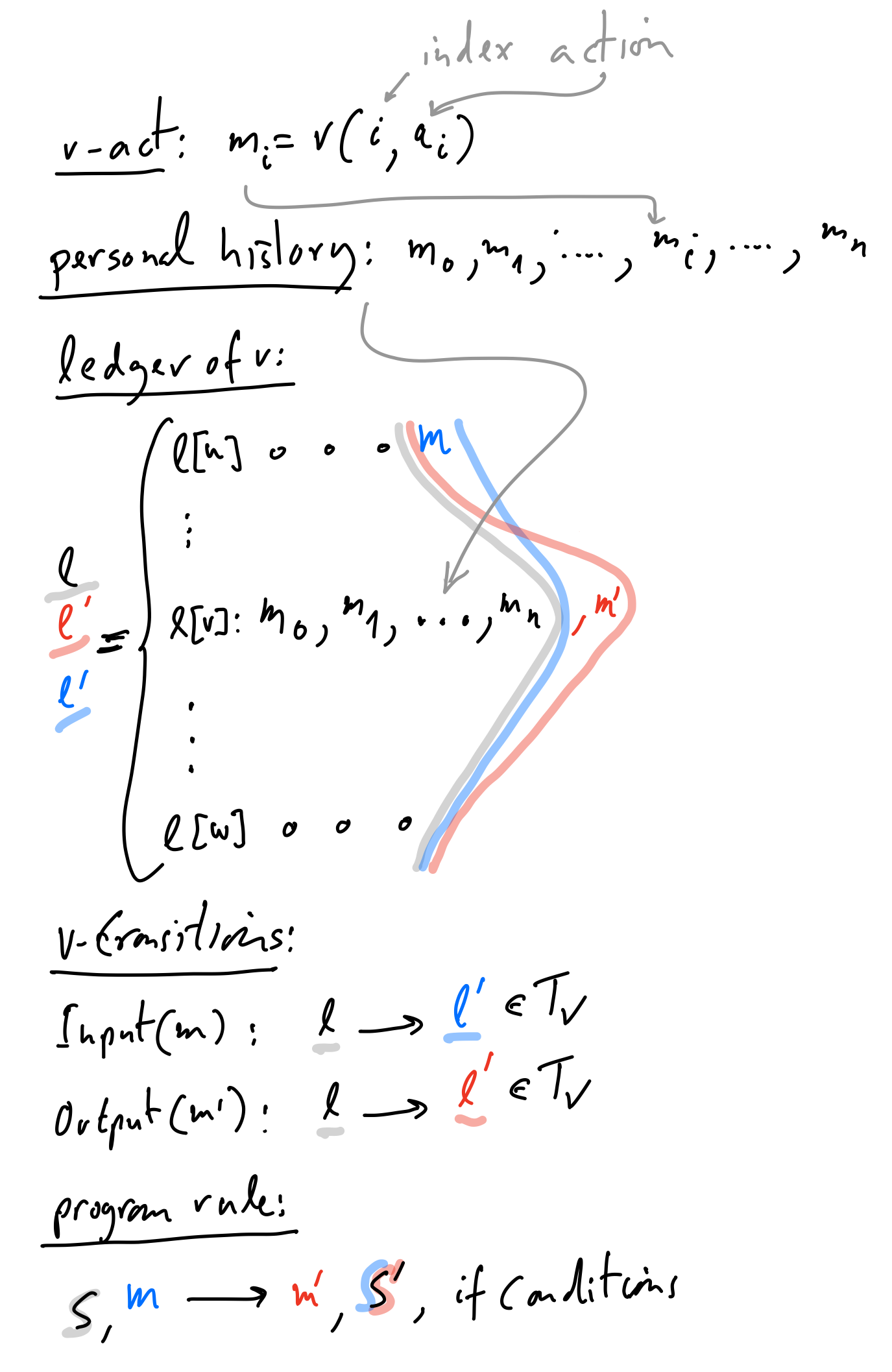}
   \caption{Agent States, Transitions and Programs: A $v$-act $m_i$, a history, a ledger $l$, an input transition (blue) inputs the $u$-act $m$, adding it to the history $l[u]$ of $u$ in $l$;  an output transition (red) that outputs the $v$-act $m'$, adding it to the history $l[v]$ of $v$ in $l$, resulting in $l'$, changes the ledger of $v$ from $l$ (to left of grey line) to $l'$ (to left of blue or red line, respectively). The \textbf{Input} transition followed by the \textbf{Output} transition can be abbreviated in a social contract program by a rule $S, m \xrightarrow{} m', S'$, in which $S$ summarizes $l$ and $S'$ summarizes $l'$ (blue and red combined).\ouri{doesn't the above rule implies that v went through the transition 'Input(m)', before initiating the transition 'output(m')'? If so, the red line in the diagram should pass to the right of message m}}
   \label{figure:elements}
\end{figure}
}

\section{A Social Contracts Programming Language}

\subsection{An Agent as a State-Transducer}

Here we outline the design of a simple social contract programming language. 
Abstractly, all that an agent can do, as a party to a social contract, is take acts based on its history, and receive acts by others.  In a practical social contract, the acts an agent can take typically depend on salient features of its history, rather than on the history in its entirety.  For example, I can pay Sue three blue coins depending on whether, according to my entire history, my current balance has at least three blue coins.  But, in this case, instead of maintaining my entire transaction history, it is enough that I maintain my balance in order to know whether I can or cannot pay Sue tree blue coins.  Similarly, I can confirm Joe's reservation depending on whether, according to my entire history, I have confirmed a reservation that is still outstanding and conflicts with Joe's reservation. Similarly, instead of maintaining my entire transaction history, it is enough that I maintain a list of outstanding reservations that I have confirmed in order to know whether I can or cannot confirm Joe's reservation.

More formally, given that social contracts employ a closed set of transitions, the possible behaviors of an agent can be characterized by a (not necessarily finite and possibly nondeterministic) state transducer.  A state transducer reads an input tape (in our case the inputs of an agent $v$) and writes an output tape (in our case the sequence of $v$-acts), changing its state in the process. 
The history of an agent fully specifies, even synchronizes, both tapes.

Hence, as an alternative to maintaining entire histories, our programming language endows each agent with an internal state, which can be viewed as a digest of its history.  As in a state transducer, the agent's state may change as a result of an input or an output.
Therefore, our proposed \emph{Social-Contracts Programming Language} (SCPL)
presents the program of an agent $v$ as a set of rules of the following form:
$$S, m \xrightarrow{} m', S',~ \emph{where Conditions}.$$
where $S$ is the internal state (state, for short) of $v$ prior to taking the transition, $m$ is an input $u$-act by some $u \ne v \in V$, $m'$ is an output $v$-act,  $S'$ is the updated state of $v$, and \emph{Conditions} are any conditions on $S$, $m$, $m'$, and $S'$, that are required for the transition to take place. The intended meaning of such a rule is that an agent in internal state $S$, upon receiving an act $m$, may take act $m'$ and change its internal state to $S'$, if \emph{Conditions} are satisfied. Note that degenerate rules, i.e., rules that do not have input $m$ and/or output $m'$ and/or \emph{Conditions} are allowed.  

In general, a programming language aims to provide a finite description to infinitely many computations; in particular, we wish our social contracts programs to describe potentially infinitely-many social contract transitions.
To do so, our programming language rules must be parameterized, namely must have variables.    SCPL incorporates logic variables and compound terms, very much in the style of logic and concurrent logic programming languages.
Next we defined the syntax of SCPL; its operational semantics via a transition system; and show that SCPL transition systems implements $\calS\calC$.

\subsection{SCPL Programs: Syntax with Examples}

A programming language needs a mechanism for binding parameters to values.  We follow concurrent logic programming languages in using matching (one-way unification) for parameter passing, and logic terms for structured data types.

\begin{definition}[SCPL Syntax]\label{definition:syntax}
These are the syntactic building blocks of SCPL programs:
\begin{itemize}
    \item The set $\calv$ of \emph{variable names} (\emph{variables}, for short) consists of all alphanumeric strings beginning with an uppercase letter, e.g. X, X1, Xs, and Foo, as well as alphanumeric strings beginning with underscore $\_$, e.g. $\_$boy; an underscore ``$\_$'' on its own is referred to as an \emph{anonymous variable}, and is a shorthand for a unique variable name not occurring anywhere else. 
    \item The set of \emph{constant names} (\emph{names}, for short) consists of all alphanumeric strings beginning with a lowercase letter, e.g. a, a1, and foo, as well as quoted strings, e.g. ``,''.
    \item The set of \emph{numbers} consists of all numeric strings, which may include a decimal point, e.g. 0, 1, 103.65.
    \item The set $\calt$ of \emph{terms} consists of all variables,  numbers, and \emph{n-ary terms} of the form $T=f(T_1,T_2,\ldots,T_n)$, $n \ge 0$,where $f$ is a name, referred to as the \emph{functor} of $T$, and each $T_i$ is a term, $i \in [n]$, referred to as a \emph{subterm} of $T$.
    \item By convention the constant name [] (read ``nil'') represents an empty list,  the binary term $[X|Xs]$ represents a list with the first element $X$ and rest $Xs$, the term $[X]$ is a shorthand for the singleton list $[X|[]]$ and the term $[X1,X2,\ldots Xn]$ is a shorthand for the nested  term $$[X1|[X2|[X3|\ldots [Xn|[]]\ldots]]].$$ 
    Note that there are infinitely many lists, and in general infinitely many terms.
    \item The set $\calR$ of \emph{social contract rules} consists of all terms of the form 
     $$\text{Output: }S\xrightarrow{} m', S'.$$
     $$\text{Input: }S, m \xrightarrow{} S'.$$
     (the full syntax of the output rule being  $``\xrightarrow{}"(S, ``,"(m',S'))$.) 
     Here is an examples of an output rule and an input rule:
\begin{verbatim}
tourist(roaming) --> 
    reserve(Host), tourist(waiting(Host)).
    
tourist(waiting(Host), Host(reservation_confirmed(Self)) --> 
    tourist(lodging(Host)).
\end{verbatim}
     In such rules, $S$,  referred to as the \emph{pre-state},  $S'$, referred to as the \emph{post-state}, and $m'$, referred to as the \emph{output act} of the rule are any terms, and $m$, referred to as the \emph{input act} of the rule,  is a unary term of the form $v(A)$ for some name $v$ and some term $A$. For programming convenience, we also provide for combined rules:
      $$S, m \xrightarrow{} m', S'.$$
      Here is an example of a combined rule:
      \begin{verbatim}
host(free), Tourist(reserve(Self))) --> 
    reservation_confirmed(Tourist), host(reserved(Tourist)).
\end{verbatim}
      Such a rule is a shorthand for the input rule $S,m \xrightarrow{}  \hat{S}$ and the output rule $\hat{S} \xrightarrow{} m', S'$, where the intermediate state $\hat{S}$ contains all variables shared between the left-hand and right-hand sides of the rule and has as a functor a new constant name that does not occur in 
      the program. For example, the general rule above is a shorthand for the following two rules:
\begin{verbatim}
host(free), Tourist(reserve(Self))) --> 
    x(Tourist).
    
x(Tourist) -->
    reservation_confirmed(Tourist), host(reserved(Tourist)).
\end{verbatim}
where \verb|x| is a name not occurring anywhere else in the program.\qed
      \end{itemize}
\end{definition}

As the output transitions of a social contract transition system depend only on the history the agent taking the transition, we wish to preserve this property in the programming language.  The following notions are instrumental in ensuring that.

\begin{definition}[Ground Term]
A term $T$ \emph{occurs} in term $T'$, denoted $T \in T'$, if $T=T'$ or if $T'$ is an $n$-ary term
$f(T_1,T_2,\ldots,T_n)$ for some constant $f$ and $T$ occurs in $T_i$ for some $i \in [n]$. A term is \emph{ground} if it no variable occurs in it, namely $X \notin T$ for every variable name $X$.  The set of \emph{ground terms} is denoted by $\calg\calt \subset \calt$.
\end{definition}

\begin{definition}[Explicit Nondeterminism]
A set of social contract rules $R$ satisfies \emph{explicit nondeterminism} if for every two ground instances of rules in $R$ with the same pre-state $S$ and with post-states $S'$ and $S''$, if $S'\ne S''$ then the two rules are not degenerate and have two different acts.
\end{definition}

Here are examples of pair of rules violating explicit nondeterminism.  
\begin{verbatim}
tourist(roaming) --> 
    reserve(Host), tourist(waiting(Host)).
tourist(roaming) --> 
    reserve(Host), tourist(lets_try_another(Host)).
    
tourist(waiting(Host1,Host2), Host1(reservation_confirmed(Self)) --> 
    tourist(lodging(Host1)).
tourist(waiting(Host1,Host2), Host1(reservation_confirmed(Self)) --> 
    tourist(waiting(Host2)).  
    
host(free) --> tourist.
host(free), Tourist(reserve(Self))) --> 
    host(confirm_reservation(Tourist)).
\end{verbatim}

While program rules have in general infinitely many instances, a program can be determined to be explicitly nondeterministic effectively using the following observation:
\begin{observation}[Conditions for Explicit Nondeterminism]
Let $R$ be a set of SCPL rules.  Then $R$ has explicit deterministic iff it has no two rules with pre-states
$S_1$ and $S_2$,  post-states $S'_1$, and acts $m_1$ and $m_2$ such that: 
$$mgu((S_1,m_1),(S_2,m_2))=\theta \text{ and } S'_1\theta \ne S'_2\theta$$
\end{observation}
\begin{proof}[Proof Sketch]
If the there are two rules violating the right-hand side of the ``iff'' then instantiate $S'_1\theta$ and $S'_2\theta$ differently thus providing a counterexample to explicit nondeterminism.  If there is a counterexample to explicit nondeterminism, then the two rules of which the counterexample is an instance also provide a counterexample to the right-hand side, as the two rule instances determine a unifier $\theta$ for which $S'_1\theta$ and $S'_2\theta$ are not identical.
\end{proof}

\begin{definition}[SCPL Role, Program]
A \emph{social contract role program} is a list of social contract rules with initial states having the same functor, referred to as a \emph{role name} (\emph{role}, for short), one of which having a 0-ary pre-state. 
An \emph{SCPL programs} is a pair $P=(P_0,R)$, where $R$ is a set of explicitly-nondeterministic social contract role programs and $P_0$, referred to as the \emph{activation of $P$}, is a list of pairs  $v\#r$, where $v$ is an agent's name and $r$ is term with its functor being a role name in $R$. The set of agent names in $P_0$ is denoted by $V(P)$, and $\calP$ denotes the set of all social contract programs.
\end{definition}

\begin{remark}[Pragmatics]
The syntax of role programs employs the following defaults and abbreviations:
\begin{enumerate}
    \item \verb|Self| is a reserved variable name, bound to the name of the agent $v$ occupying the role.
    \item A $v$-act $m=v(a)$ is signed by $v=\verb|Self|$ implicitly, so the output act $a$ is perceived by all agents as the signed act \verb|Self(|$a$\verb|)|.
    \item Input acts that have no rules for handling them are ignored, namely received without changing the agent's state.  Typically, these would be acts relevant to other agents.
\end{enumerate}
\end{remark}

\subsection{An Online Lodging Marketplace}

Here we present a distributed social contract among tourists and hosts.  It is highly simplified, just to illustrate the concept:
\begin{enumerate}
    \item There is a bunch of tourists and a bunch of hosts; initially all hosts are free.
    \item A tourist may request a room from a host and wait; if the room is free, then the host grants the request and records the room as booked; if the room is booked, then the host denies the request.
    \item If the request is granted, then the tourist checks in, and then checks out; when the tourist checks out, the host records that the room is free.  If the request is denied, then the tourist may request a room again, from the same host or from another host.
    \item No money exchange here.
\end{enumerate}

First, the host:
  The state of the host can be either \verb|free| or \verb|reserved(Tourist)|.   If an \verb|free| host received a \verb|reserve(Self,Tourist)| request, it grants it, by stating \verb|granted(Tourist)| and changing its state to \verb|reserved(Tourist)|. When the tourist that reserved the room checks out, the booked owner becomes \verb|free| again.  

The following are SCPL roles for a simple-minded Airbnb-less social contract among hosts and tourists.
First, an example of a \verb|host| role program:

\Program{Host Role}\label{program:host}
\begin{verbatim}
host --> host(free).

host(free), Tourist(reserve(Self))) --> 
    reservation_confirmed(Tourist), host(reserved(Tourist)).
host(reserved(Tourist), Tourist1(reserve(Self))) --> 
    reservation_denied(Tourist1), host(reserved(Tourist))
    where Tourist =\= Tourist1.
host(reserved(Tourist)), Tourist(checkout(Self)) --> host(free).
\end{verbatim}

Now, the tourist:
  The state of the tourist can be either \verb|roaming|, \verb|waiting| for a response to a reserve request,
or \verb|lodging(Host)|.
  
The following is an example of a \verb|tourist| role program:

\Program{Tourist Role}\label{program:tourist}
\begin{verbatim}
tourist --> tourist(roaming).

tourist(roaming) --> 
    reserve(Host), tourist(waiting(Host)).
    
tourist(waiting(Host), Host(reservation_confirmed(Self)) --> 
    tourist(lodging(Host)).
    
tourist(waiting(Host)), Host(reservation_denied(Self)) --> 
    tourist(roaming).
   
tourist(lodging(Host)) --> 
    checkout(Host), tourist(roaming).
\end{verbatim}

Here are the rules for a tourist $v$:
  The first rule is an output rule,  missing an incoming message $m$, so this rule can apply whenever the agent's state is $S$.  Note that a \verb|tourist| nondeterministically chooses a room owner to reserve a room from. In our interpretation, this nondeterminism is realized as a query to the person operating this agent, which room would you like to reserve?  If and when the person chooses from which  \verb|Host| to reserve, the agent then requests to reserve the room, sending the message \verb|reserve(Host)|. The agent then waits for the response.  If  the reservation is \verb|reservation_confirmed| it checks in at its leisure.  If the request is \verb|reservation_denied|, it may choose again a host to reserve from.

The role programs as written can accommodate multiple tourists and multiple hosts in the same social contract.  Here is an example of an activation of a social contract with these two roles:

\Program{Activation of a Tourists/Hosts Social Contract}
\begin{verbatim}
[nimrod#host,udi#tourist,avigail#tourist,gal#tourist,ouri#host]
\end{verbatim}

Here is a possible history of executing this social contract thus activated:\footnote{This is actually a trace of a Concurrent Prolog program hand-compiled from this social contract.  The trace include also the ``Oracle choices'' of the various agents. }\\

\begin{verbatim}
H / 1 = gal(oracle(reserve(ouri)))
H / 2 = udi(oracle(reserve(nimrod)))
H / 3 = avigail(oracle(reserve(nimrod)))
H / 4 = udi(reserve(nimrod))
H / 5 = gal(reserve(ouri))
H / 6 = nimrod(reservation_confirmed(udi))
H / 7 = avigail(reserve(nimrod))
H / 8 = nimrod(reservation_denied(avigail))
H / 9 = ouri(reservation_confirmed(gal))
\end{verbatim}

We now consider a different social contract, where a broker that accumulates a list of reservations and a list of free rooms, assuming all rooms are equivalent, and grants reservations on a first-come-first-served basis.  

\Program{Brokered Hosts/Tourists Social Contract}

\begin{verbatim}
host --> host(free)
host(free) --> free, host(waiting).
host(waiting), broker(reservation(Tourist,Self)) --> host(reserved(Tourist)).
host(reserved(Tourist)), checkout(Tourist,Self) --> host(free).

tourist --> tourist(roaming).
tourist(roaming), reserve --> tourist(waiting).
tourist(waiting), broker(reservation(Self,Host)) --> tourist(lodging(Host)).
tourist(loading(Host)) --> checkout(Host), tourist(roaming).

broker --> broker([],[]).
broker(Rooms,Reservations), Host(free) -->
    broker(Rooms',Reservations) 
    where Rooms' is the result of appending 
    Host to Rooms.
broker(Rooms,Reservations), Tourist(reserve) -->
    broker(Rooms,Reservations') 
    where Reservations' is the result of
    appending Tourist to Reservations.
broker([Host|Rooms],[Tourist|Reservations]) -->
    reservation(Tourist,Host)), 
    broker(Rooms,Reservations)
\end{verbatim}

This is not far from how Airbnb essentially operates. 
With a broker, one may have two separate social contracts. One among room owners and the broker (Airnbn), and one among the tourists and the broker (Airnbn again).  The broker, then, is a party to both social contracts.  The broker may receive a confirmation signed by the room owner via the social contract of the owners, and forward it to the tourist, via the tourists social contract.  Such a signed confirmation is non-repudiated, even if obtained indirectly.  Third, one may add payments, as in real life. Realizing a currency network using digital social contracts is described in a companion paper~\cite{EgalitarianCurrencyNetworks}.  However, in our realization, the broker need not collect rent from transactions among the tourists and the room owners, as Airbnb does.  It could be owned and operated cooperatively by the tourists, by the owners, or by both. The issue of shared, autonomous agents (aka smart contracts) will be discussed in detail in a subsequent paper.  Also, in this paper we have eschewed the programming language design issue of how to bind formal parameters to actual parameters in a social contract, parameter scope, and how to start a social contract. These are standard issues of programming language design, but fixing these details in this high-level paper is premature. The following muck-up code, without formal semantics yet, may illustrate a design option, assuming all names stand for public keys, for which the private keys are known by the respective agents.

\subsection{A Egalitarian Currency}

As another example, consider an egalitarian currency~\cite{EgalitarianCurrencyNetworks}.  The social contract assumes a clock agent that issues clock ticks.  Each party to the social contract may mint one coin at each clock tick, may receive payments from other agents, and may pay other agents if its balance affords the payment.

\Program{Egalitarian Currency}
\begin{verbatim}
agent --> agent(0).

agent(Balance), clock(tick) --> 
    agent(Balance') where Balance' := Balance + 1.
    
agent(Balance), Other(pay(Self,X)) -->
    agent(Balance') where Balance' := Balance + X.
   
agent(Balance) --> 
    pay(Other,X), agent(Balance') 
    where Balance >= X & Balance' := Balance - X.
\end{verbatim}
Note that in the last rule of \verb|agent|, both the addressee of the payment, \verb|Other|, and the amount \verb|X|, are free variables, to be determined by volition of the payer.  
Similarly, which \verb|Host| to reserve from, is a free variable in the \verb|Tourist| program, to be determined by volition of the tourist.  As discussed above,  the intention is that in the implementation of the Social Contracts Programming Language, a computational agent will take volitional acts at the advice of the person operating the agent.\footnote{Of course, a person may operate computational agents, possibly endowed with Artificial Intelligence, that undertake volitional acts on her behalf; investigating this direction is beyond the scope of this paper.} An exception may be a clock agent, which will use a physical clock (or in any case an outside computational agent connected, indirectly, to a physical clock) to decide when to tick.  Formally, and in the implementation of the Social Contracts Programming Language, we view the external agent as an Oracle that resolves explicit output nondeterministic choices enabled by the contract.

\subsection{A Simple Data Storage Service}

A similar example of rental goods is a distributed data storage (cloud) service.   Here is a simplified setting:

\begin{enumerate}
    \item There are a bunch of users and a bunch of data storage owners; initially all storage spaces are free.
    \item A user may request some storage space (to store his data) from an owner and wait; if some space is free, then the owner grants the request, possibly partially, and records the granted space as booked; if the space is booked, then the owner denies the request.
    \item If the request is granted, then the user stores her data, and then checks out; when the user checks out, the owner records that the space is free.  If the request is denied, then the user may request a space again, from the same owner or from another owner.
    \item No money exchange here.
\end{enumerate}

The social contract for this setting is similar to the unbrokered hosts/tourists setting, with the addition that a reservation requests capacity, and can be satisfied fully or partially.  In the brokered setting, there is an added complication that requests can be satisfied in fractions, so leftover storage factions for both users and hosts should be managed.

\subsection{Citizens Band and Social Community Examples}

Here we consider dynamically growing a social contract, in a simple ``Citizens Band'' (CB) application, where anyone can talk, everyone hears what everyone else says, and anyone may invite anyone to join.

\Program{Citizens Band}
\begin{verbatim}
agent -->  say(X), agent.
agent -->  Friend#agent, agent.

agent, Foo(hi_there) --> welcome(Foo), agent.
\end{verbatim}

We now show the social contract of a social community with managers and members.
Only a manager can add members or turn members to managers.

\Program{WhatsApp-like Group with a Manager}
\begin{verbatim}
manager -->  manager([]).
    
manager(Members) -->
    Friend#member, manager([Friend|Members]).
manager(Members) -->
    please_leave(Member), manager(Members') 
    where Members' is a result of removing Member from Members.
manager([]) --> stop.

member --> says(X), member.
member, _(please_leave(Self)) --> says(bye), stop.
member --> says(bye), stop.
\end{verbatim}

\subsection{Egalitarian Governance of Social Contracts}\label{section:democracy}

The examples up to this point allow agents to realize sovereignty, transparency, and privacy over a shared medium of interaction. Here we consider the standard principle of democracy, generally abbreviated as ``one person, one vote''. That is, we show how a community may form democratically.  This method of democratic community formation could be applied to any of the social contracts mentioned above - the Social Community, Hosts and Tourists, and Egalitarian Currency.  Here decisions to add or remove a member are taken by a simple majority, but extending the contract to require a supermajority for taking a decision is not difficult.

The democratic community has a secretary that handles the ballot.
The secretary is deterministic, and as such does not need an Oracle.
Hence, it is autonomous and not bound to any specific individual.  It is initiated by the founder.

\Program{Democratic WhatsApp-like Group}
\begin{verbatim}

founder -->  autonomous#secretary([Self]), member.

secretary -->  secretary([]).
secretary(Members), _(propose(X)) -->
    ballot(X,Members,0), secretary(Members).
secretary(Members), ballot(X,[],Result) -->
    secretary_apply(X,Result,Members). 
    
secretary_apply(X,Result,Members) :-
    secretary(Members) where Result =< 0.
secretary_apply(add(Member),Result,Members) :-
    Member#member, secretary([Member|Members]) 
    where Result > 0.
secretary_apply(remove(Member),Result,Members) :-
    please_leave(Member), secretary(Members') 
    where Result > 0, 
    Members' is the result of removing Member from Members.

member --> says(X), member.
member, ballot(X,[Self|Members],R) --> 
    ballot(X,Members,R'), member
    where R' := R, R+1, or R-1.
member, _(please_leave(Self)) --> says(bye), stop.
member --> says(bye), stop.
\end{verbatim}

Note that this is a very simple program, in particular as (1) there is no option to decline an invitation (2) agents must wait for other agents to vote. Indeed, we provide the program mainly as a proof of concept, but to implement a practical democratic decision process, more complex program is needed.  We also note that votes are public to the community members and are not anonymous.

\section{Operational Semantics of SCPL}

The operational semantics of SCPL is given by mapping every SCPL program $P=(P_0,R)$ into a transition system $SCDS$ (for Social Contract with a Distributed State). 
We first define the model, then show how SCPL programs are mapped into SCDS transition systems, and finally show that SCDS implement SC.

We assume three given sets: $V$ of agents,  $A$ of agent actions,  $\cals$ of agent states.

\begin{definition}[Transition Function]
We assume transition function $\tau: V \times \cals \times \calM \xrightarrow{} \cals \cup \{\bot\}$, where $\tau(v,s,u(a))$ is referred to as an \emph{output transition} if $u=v$ and \emph{input transition} if $u\ne v$. An input transition is always defined, namely different from $\bot$: $\tau(v,s,u(a)) \in S$ for every $u \ne v \in V$, $s \in S$, and $a \in A$.
\end{definition}

\begin{definition}[Addressed Message Store]
A set $M \subseteq \calA\calM := V \times \calM$ is referred to as an \emph{addressed message store}.  A $v$-act $m=v(a)$ is sent to $u$ through $M$ by adding $(u,v(a))$ to $M$, and is received by $u$ removing it from $M$. 
\end{definition}

\begin{remark}[Signed Messages, Ordered Messages]
A message  $(u, m)$ in the message store indicates that some $v$ ha sent the message $m$ to $u$, and $u$ has not received it yet. Recall that the agent $v$ signs the message $m$, so, even though we store objects of the form $(u, m)$, where $u$ is the destination and $m$ is the message, the original sender $v$ is implicitly there as well, as the signatory of $m$.
Furthermore, peer-to-peer messages need to be processed by the recipient in the order sent.  To achieve that, we could  number actions sequentially and order them upon receipt.  Instead, we simply assume that the message store delivers point-to-point messages in the order sent.  We relax this assumption in a companion paper, when discussing fault-tolerant implementations~\cite{ftdsc}.
\end{remark}

\begin{definition}[Distributed State Configurations]
The set of \emph{SCDS configurations} $\calC \subset \cals^V \times \calM$  consists of all pairs $c=(l,M)$ where $l \in \cals^V$ is agents' state and $M \in \calM$ is a message store.
\end{definition}

\begin{definition}[SCDS transitions]
The set of \emph{SCDS transitions} $T(\tau)$ over $\calC$  consist of all pairs
$c \xrightarrow{} c' \in \calC \times \calC$, where $c =(l,M)$, $c' = (l',M')$,  $c' = c$ except for $l'_v:=\tau(v,l_v,u(a))\ne \bot$, $u, v\in V$, $a \in A$ and either:
\begin{enumerate}
    \item Output: $u=v$, $M' = M \cup \{(w, v(a)) : w \neq v \in V\}$.
    \item Input:  $u \ne v$, $(v, u(a))\in M$ and $M' = M \setminus \{(v, u(a))\}$. 
\end{enumerate}
In which case the transition is referred to as an \emph{output $v$-transition} and can also be written as $c \xrightarrow{(v,v(a))} c'$, or \emph{input $v$-transition} and can also be written as $c \xrightarrow{(v,u(a))} c'$, respectively.\qed
\end{definition}

\begin{definition}[SCDS Transition System]
An SCDS transition system $(S,s_0,T)$ has configurations $S \subset \cals^V \times \calM$ and transitions $T$ defined via 
agents $V$, actions $A$, and a transition function $\tau: V \times S \times \calM \xrightarrow{} S \cup \{\bot\}$, to be $T := T(\tau)$.
\end{definition}

\begin{proposition}[History Determines State]\label{proposition-hds}
In an $SCDS$ run, the history of an agent uniquely determines its state.
\end{proposition}
\begin{proof}[Proof Sketch]
By induction on history length.  The initial state of every agent $v$ is uniquely determined by $s_0$. 
Thanks to $\tau_P$, current agent state is mapped by the next action to a unique agent state.
\end{proof}

We specify now the configurations and transition function for an SCPL program $P$: The set of actions $A$ is the set of all ground terms, $A := \calg\calt$,  the set of agents is $V := V(P)$ and the set of agent's states is the set of all  ground terms $\cals := \calg\calt$.

\begin{definition}[Distributed State Corresponding to a Program]
Given an SCPL program $P=(P_0,R)$, the \emph{corresponding set of SCDS configurations} $\calC(P) \subset \cals^V \times \calM$  consists of all pairs $c=(s,M)$ where $s \in \cals^V$ is agents' state and $M \in \calM$ is a message store, and the initial state $s_0$ determined by $P_0$.
\end{definition}

We now derive SCDS transitions from SCPL programs. Each SCPL rule with variables stands for infinitely-many SCDS transitions, one for each instance of the rule, as defined next.  
Recall that a substitution $\theta: \calv \xrightarrow{} \calt$ is a mapping from variables to terms; for any substitution $\theta$ and term $T$,  denote by $T\theta$ the result of replacing every variable $X \in T$ by $\theta(X)$, and refer to $T\theta$ as an \emph{instance} of $T$;  and as a \emph{ground instance} of $T$ if $T\theta$ is ground.  

\begin{definition}[Grounded Program]\label{definition:grounded}
Given a program $P=(P_0,R)$, we define its \emph{grounded roles} $R^*$ in several steps,  as follows:
\begin{enumerate}
    \item  \textbf{Signatures}: Modify every output rule in $R$ by adding \verb|Self(_)| to the output action.  For example, the first clause of the \verb|tourist| role above becomes:
\begin{verbatim}
tourist(roaming) --> 
    Self(reserve(Host)), tourist(waiting(Host)).
\end{verbatim} 
    \item  \textbf{Ground Instances}: Let the set of grounded rules $R^*$ include all ground instances of any  input rule  $S,m \xrightarrow{} S' \in R$ and output rule  $S \xrightarrow{} m', S' \in R$.    An example ground instance of the rule above is:
    \begin{verbatim}
tourist(roaming) --> 
    luca(reserve(nimrod)), tourist(waiting(nimrod)).
\end{verbatim} 
    \item  \textbf{Closure for Inputs}: Add to $R^*$ all ground input rules $S,m \xrightarrow{} S$ for which $R^*$ has no ground rule $S,m \xrightarrow{} S'$, where $S \ne S'$.
\end{enumerate}
\end{definition}
Note that since the set of terms is infinite, if $P$ is not ground then $R^*$ is infinite.

\begin{definition}[SCPL transition function]
With each program $P=(P_0,R)$ we associate a state-transition function $\tau_P:  V \times \cals \times A \xrightarrow{} \cals$, defined for every $u \ne v \in V$, $S \in \calS$, and $a \in A$ as follows:
\begin{enumerate}
    \item $\tau_P(v,S,u(a)) := S'$ if $S,u(a) \xrightarrow{} S' \in R^*$, $\bot$  otherwise.
    \item $\tau_P(v,S,v(a)) := S'$ if $S \xrightarrow{} v(a) , S' \in R^*$.
\end{enumerate}
\end{definition}
\begin{observation}
The transition function $\tau_P$ is well-defined.
\end{observation}
\begin{proof}
Thanks to explicit output-nondeterminism, case 1 is unique if defined.
Thanks to input-determinism and input closure,  case 2 is well defined.
\end{proof}

\begin{definition}[SCDS transitions Corresponding to a Program]
Given an SCPL program $P=(P_0,R)$, the set of \emph{corresponding SCDS transitions} $T(P)$ over $\calC(P)$  consist of all pairs
$c \xrightarrow{} c' \in \calC(P) \times \calC(P)$, where $c =(s,M)$, $c' = (s',M')$,  $c' = c$ except for $s'_v:=\tau_P(v,s_v,u(a))\ne \bot$, $u, v\in V$, $a \in A$ and either:
\begin{enumerate}
    \item Output: $u=v$, $M' = M \cup \{(w, v(a)) : w \neq v \in V\}$.
    \item Input:  $u \ne v$, $(v, u(a))\in M$ and $M' = M \setminus \{(v, u(a))\}$. 
\end{enumerate}
In which case the transition is referred to as an \emph{output $v$-transition} and can also be written as $c \xrightarrow{(v,v(a))} c'$, or \emph{input $v$-transition} and can also be written as $c \xrightarrow{(v,u(a))} c'$, respectively.\qed
\end{definition}

\begin{definition}[SCDS Transition System Corresponding to a Program]
Given an SCPL program $P=(P_0,R)$, its \emph{corresponding transition system} $SCDS(P) = (\calC(P),T(P))$ has configurations $\calC(P)  \subset \cals^V \times \calM$ as states and transitions $T(P) \subseteq \calC(P) \times \calC(P)$.  The family of SCLP transitions systems is denoted by $SCDS(\calP) := \{SCDS(P) ~|~ P \in \calP)$.
\end{definition}

We show that $\calS\calC\calS\calS$  implements $\calS\calC$ by describing a compiler that takes an abstract SC transition system and returns a  state-based distributed transition system SCDS that implements SC.  It does so by first compiling SC to an SCPL program, and then map the program to its operational semantics SCDS. We gain here not only the understanding that SCDS indeed implements SC, but also the sense in which social contracts described via SC are in fact specifications for SCPL programs.

\begin{definition}[Compiling SC to SCDS]
We define $\emph{AtoD}: \calS\calC \xrightarrow{} \calS\calC\calD\calS$ as follows.
Given an $\calS\calC$ transition system $SC=(S,l_0,T)$, we first map it to the grounded SCPL program $P=(P_0,R^*)$ where $P_0= (\Lambda^V,\emptyset)$ and for every $l \xrightarrow{(v,u(a))}l' \in T$:
\begin{enumerate}
    \item Input:  $l_v, u(a) \xrightarrow{} l'_v \in R^*$ if $u=v$
    \item Output: $l_v \xrightarrow{} u(a), l'_v \in R^*$ if $u\ne v$
\end{enumerate}
We then define  $\emph{AtoD}(SC):= SCDS(P)$.
\end{definition}
We note in passing that $P$ could be, in principle, infinite.  However, if bisimilarity induces a finite  number of equivalence classes on $S$, then instead of the post-state of a transition being $l'_v$, it could be a short representative of the equivalence class of $l'_v$, and by looping thus, $P$ could also be finite.

\begin{proposition}
\emph{AtoD} is a compiler.
\end{proposition}
To prove this proposition, we have to show that $\emph{AtoD}(SC)$ implements $SC$, and to do so we have to show a mapping from $SCDS$ states to $SC$ states and prove it to be an implementation of $SC$ by $SCDS:=\emph{AtoD}(SC)$.
Recall that for SC output $v$-transitions, the closure condition ensures that the  history of $v$ is sufficient to determine what transitions $v$ can take, and under the $\emph{AtoD}$ compiler, this history is indeed encoded in $v$'s state $s_v$.  For SC input transitions, what matters is the difference between the history of $v$, $l_v$, and the diagonal $l^*$, and a sound input transition can only take as input an action of some other agent $u$ that is next in difference between the $u$-acts in the history of $v$, $l_v[u]$, and the diagonal $l^*_u=l_u[u]$.
Hence, we show that in  $\emph{AtoD}(SC)$ computations, the message store captures precisely the difference between the diagonal and the histories of agents, and therefore provides a sound basis for sound input transitions.

\begin{observation}[State determines Message Store]\label{observation:ltom}
Let SC be a social contract and $SDSC = \emph{AtoD}(SC)$.  Consider a computation
$c_0 \xrightarrow{} c_1 \xrightarrow{}$, where $c_i =(s^i,M_i)$.  Then $M_i$ contains exactly all messages $(v,u(a))$, such that $a \in s^{i*}_u \setminus s^i_v$, for all $u\ne v \in V$.
\end{observation}
\begin{proof}
We prove by induction and a simple case analysis. The claim holds vacuously for $n=0$.  Assume it holds for $n>0$ and consider $c_n \xrightarrow{(v,u(a))}c_{n+1}$.  There are two cases:
\begin{enumerate}
    \item Input:  $v\ne u$, in which case following the transition, the message $(v,u(a))$ is removed from $M$ and added to $s^{n+1}_v$, preserving the inductive assumption.
    \item Output:  $v=u$, in which case following the transition, the message $(w,v(a))$ is added to $M$ and added $M)_{n+1}$, preserving the inductive assumption.
\end{enumerate}
\end{proof}

\begin{lemma}
$\emph{AtoD}(SC)$ implements $SC$.
\end{lemma}
\begin{proof}
First we note that with this compiler, $\cals = \calH$ and $\cals^V = \calL$ (up to syntactic variations).
Consider the mapping $f: \cals^V \times \calM \xrightarrow{} \cals^V = \calL$, where $f(s,M) := s$. Namely, the mapping ignores the messages store.  We claim that $f$ is a strict morphism.  In an output transitions, the state of each agent is precisely its history, so there is no difference between $SC$ and $\emph{AtoD}(SC)$ in that regard, and hence.  Regarding input  transitions, as noted in Observation \ref{observation:ltom}, the message store captures precisely the difference between the ledger and the diagonal, and therefore each SDSC  transition corresponds precisely to an SC transition under $f$. In other words, $f$ is a strict morphism. 
\end{proof}

\begin{corollary}
$\calS\calC\calD\calS$ implements $\calS\calC$.
\end{corollary}
\begin{proof}
The compiler $AtoD$ constitutes a constructive proof.
\end{proof}

\section{Outlook}

We have introduced the concept of digital social contracts,  provided a mathematical definition of them,
outlined a design for a social contracts programming language, and demonstrated its social utility via program examples.  Much remains to be done; some is discussed in companion papers~\cite{ftdsc,EgalitarianCurrencyNetworks}.

\section*{Acknowledgements}
Ehud Shapiro is the Incumbent of The Harry Weinrebe Professorial Chair of Computer Science and Biology.
We thank the generous support of the Braginsky Center for the Interface between Science and the Humanities.
Nimrod Talmon was supported by the Israel Science Foundation (ISF; Grant No. 630/19).

\bibliographystyle{plain}
\bibliography{bib}

\end{document}

%% file: groupcontract.tex


\begin{flushleft}
\begin{table}
%
%

 \begin{tabular}{ |  m{4em}|| m{16em} | } 
 \hline
  \textbf{Role} & \textbf{Actions}  \\ [2ex] 
 \hline\hline 
  \textbf{None} & 
\begin{enumerate}
    \item Initiate a community as a manager
    \item Join a community as an member
\end{enumerate}
 \\ [2ex]
 \hline
 \textbf{Manager} & 
\begin{enumerate}
    \item Invite/remove a member
    \item Do anything a member can do
\end{enumerate}
  \\ [2ex]
 \hline
 \textbf{Member} & 
\begin{enumerate}
    \item See the manager and members
    \item See messages sent by others
    \item Send a message 
    \item Leave the community
\end{enumerate}
  \\ [2ex]
 \hline
\end{tabular}
\caption{The \textbf{Community} Social Contract.} 
\label{figure:community}
\end{table}
\end{flushleft}